\documentclass[fleqn]{2026_arxiv}
\usepackage{bm}
\usepackage[T1]{fontenc}
\usepackage[utf8]{inputenc}

\usepackage{xcolor}

\begin{document}
\ensubject{subject}
\ArticleType{Article}
\Year{2026}
\Month{xx}
\No{x}
\DOI{??}

\title{Simultaneous estimation of relative phase and coherence in astronomical interferometry}


\author[1]{Yawen Tang}{}%
\author[2]{Wei Ye}{}%
\author[1]{Lu Qin}{}%
\author[1]{Jinxin Li}{}
\author[1]{Xinxin Wang}{}%
\author[1]{Zunlue Zhu}{}
\author[1]{\\Shoukang Chang}{\textsuperscript{*}}%
\author[3]{Shao-Ming Fei}{\textsuperscript{*}}%
\author[1]{Xingdong Zhao}{\textsuperscript{*}}%

\AuthorMark{Yawen Tang}


\address[1]{School of Physics, Henan Normal University, Xinxiang 453007, China;}
\address[2]{School of Information Engineering, Nanchang Hangkong University, Nanchang 330063, China;}
\address[3]{School of Mathematical Sciences, Capital Normal University, Beijing 100048, China}


\abstract{Astronomical interferometry is a cornerstone technique for high-resolution stellar imaging 
and observational astrophysics, extracting spatial information from the coherence of light collected by 
separated telescopes. Since the degree of coherence is complex, a genuine imaging task requires 
the joint recovery of the modulus and the relative phase, instead of independent single-parameter estimations. 
We investigate the simultaneous estimation of both parameters based on direct interferometry scheme and 
continuous-variable quantum teleportation scheme. We find that in simultaneous estimation the direct interferometry scheme consistently yields
a lower quantum Cramér-Rao bound, demonstrating its superiority over the continuous-variable quantum teleportation scheme. 
Furthermore, we establish the conditions under which the classical Cramér-Rao bound for Gaussian measurements saturates 
the quantum Cramér-Rao bound, identifying heterodyne detection as a near-optimal measurement scheme 
in the large mean photon number regime. An analysis of transmission loss reveals that the direct interferometry scheme 
yields superior precision in the short-baseline regime, whereas the continuous-variable quantum teleportation scheme outperforms it at longer baselines.}


\maketitle
\begingroup
\renewcommand{\thefootnote}{\fnsymbol{footnote}}

\makeatletter
\renewcommand{\@makefnmark}{\hbox{\textsuperscript{\@thefnmark}}}
\renewcommand{\@makefntext}[1]{%
  \noindent\@makefnmark\,#1%
}
\makeatother

\footnotetext[1]{%
Corresponding authors
(Shoukang Chang, email: \href{mailto:changshoukang@htu.edu.cn}{changshoukang@htu.edu.cn};\\
Shao-Ming Fei, email: \href{mailto:feishm@cnu.edu.cn} {feishm@cnu.edu.cn};
Xingdong Zhao, email:\\  \href{mailto:phyzhxd@gmail.com}{phyzhxd@gmail.com})
}
\endgroup


\begin{multicols}{2}
\section{Introduction}\label{section1}
Astronomical interferometry has revolutionized observational astronomy by
providing a powerful framework to overcome the diffraction limit of single
telescope apertures \cite{1,2,3}. By synthesizing optical signals captured
by spatially separated apertures, an interferometer effectively mimics a
giant synthetic telescope, enabling the extraction of fine spatial details
of distant stellar objects at milliarcsecond or microarcsecond scales \cite%
{4,5,6,7}. According to the seminal van Cittert-Zernike theorem, the spatial
coherence function of the radiation field emitted by a distant, incoherent
astronomical source encodes its vital geometric and structural properties
\cite{8}. Specifically, the relative phase and the degree of coherence
between different spatial points characterize the source's angular position
and spatial profile, respectively \cite{9,10,11,12}. In direct
interferometry schemes, these quantities are measured by physically routing
the collected stellar photons through optical channels to a central station
where they undergo interference \cite{9,13,14,15}. However, as the baseline
is extended to achieve higher resolution, the transmission loss within these
physical transport channels scales exponentially with the distance \cite%
{16,17,18}. For optical frequencies, this severe attenuation drastically
diminishes the signal-to-noise ratio, imposing a fundamental physical
bottleneck on the maximum achievable baseline length and capping the
resolution of classical optical arrays \cite{19,20,21}.

To circumvent this seemingly insurmountable limitation imposed by
transmission loss, the paradigm of quantum enhanced telescopy has emerged,
leveraging the principles of quantum information science to revolutionize
astronomical imaging \cite{13,22,23}. The seminal proposal by Gottesman,
Jennewein, and Croke demonstrated that predistributed quantum entanglement
could serve as a nonlocal resource to bypass the direct transmission of
fragile stellar photons \cite{13}. In the standard framework proposed by Gottesman, Jennewein, and Croke and its discrete-variable extensions, an auxiliary single photon is coherently delocalized over two remote telescope modes and shared between the two telescope stations\cite{13}. A joint quantum measurement performed locally on the
incoming stellar photon and the ancillary entangled photon effectively
teleports the quantum state of the starlight to a central processor without
suffering from transmission loss that is dependent on the baseline \cite{13}%
. Subsequent theoretical developments have enriched this field by exploring
multinode quantum assisted telescope networks, incorporating strict physical
constraints such as developing random distillation protocols to optimize
resource distribution, as well as proposing auxiliary quantum sources that
are independent of loss \cite{4,14,15,16}. Nevertheless, the implementation
of these discrete variable protocols imposes exceedingly stringent
experimental requirements \cite{24,25,26,27}. They fundamentally rely on
high fidelity on demand single photon sources, low noise quantum memories,
and photon number resolving detectors \cite{28,29,30}. Given the current
state of quantum technological infrastructure, scaling these discrete
architectures to realistic astronomical baselines remains challenging \cite%
{24,25,26,27}.

To overcome the infrastructural hurdles of discrete variable architectures,
continuous variable quantum teleportation has recently emerged as a highly
practical alternative \cite{24,25,26,27}. By utilizing two-mode squeezed
vacuum states as entanglement resources, continuous variable schemes can be
implemented using mature linear optics and can uniquely exploit the
multiphoton events naturally present in thermal stellar states \cite%
{24,25,26}. However, the extent to which continuous-variable quantum
teleportation provides a genuine quantum advantage remains heavily debated
\cite{24,26}. Some recent analyses indicate that continuous variable schemes
utilizing standard unconditional measurements, such as homodyne detection,
fail to outperform direct interferometry when considering cumulative Fisher
information \cite{24,26}. Similarly, evaluating the quantum Fisher
information reveals that in lossless or low loss regimes, continuous
variable teleportation generally underperforms direct interferometry,
exhibiting only a marginal advantage in specific high loss regions while
requiring experimentally prohibitive squeezing levels \cite{24,26}.
Beyond these technological considerations, a more fundamental motivation arises from the imaging nature of astronomical interferometry itself. The ultimate goal is not merely to estimate either a relative phase or a degree of coherence in isolation, but to reconstruct the spatial intensity distribution of a distant incoherent source. According to the van Cittert-Zernike theorem \cite{8}, each interferometric baseline samples a Fourier component of the source intensity distribution through the mutual coherence function, or equivalently the complex degree of coherence. This coherence quantity is intrinsically complex: its modulus characterizes the strength of coherence, while its argument gives the relative phase associated with the spatial position and structural asymmetry of the source. Therefore, estimating only one of these quantities implicitly assumes that the other is known or fixed, which does not reflect a genuine imaging task. A faithful assessment of continuous-variable quantum stellar interferometry should treat the complex degree of coherence as a multiparameter object and determine the jointly attainable precision for both modulus and phase. This is particularly important because the measurements that are optimal for the two parameters separately need not be jointly optimal, and the apparent advantage predicted by single-parameter quantum Fisher information may be reduced once the measurement incompatibility and practical measurement schemes are taken into account. Consequently, a rigorous simultaneous-estimation framework is essential for evaluating the true imaging capability and quantum advantage of continuous-variable stellar interferometry.

In this paper, we investigate the simultaneous estimation problem of the
relative phase and the degree of coherence in astronomical interferometry,
contrasting to a direct interferometry with a continuous-variable quantum
teleportation scheme. Although the symmetric logarithmic derivative
operators \cite{31} for these parameters are non-commuting in both schemes,
we show that the mean Uhlmann curvature matrix \cite{32,33,34} is a zero matrix, ensuring
that the quantum Cram\'{e}r-Rao bound (QCRB) \cite{31} is asymptotically
tight. Our findings reveal that the direct interferometry consistently
yields a lower QCRB, outperforming the continuous-variable quantum
teleportation scheme. Additionally, we determine the conditions under which
the Cram\'{e}r-Rao bound (CRB) for Gaussian measurements saturates the QCRB,
identifying heterodyne measurement as near-optimal in the large mean photon
number regime. Lastly, we analyze the effects of transmission loss, showing
that direct interferometry yields superior precision in the short-baseline
regime, whereas the continuous-variable quantum teleportation scheme
outperforms it at longer baselines.

The remainder of this paper is arranged as follows. In Sec. II, we review
the known results of Gaussian multiparameter quantum estimation theory. In
Sec. III, we explore the quantum simultaneous
estimation of the relative phase and the degree of coherence. Our main
conclusions are drawn in the last section.

\section{Preliminaries}\label{section2}

We first recall the Gaussian multiparameter
quantum estimation. Consider an $m$-mode bosonic Gaussian system
characterized by a vector of quadrature operators $\hat{r}=(\hat{x}_{1},\hat{%
p}_{1},...,\hat{x}_{m},\hat{p}_{m})^{\text{T}}.$ The canonical commutation
relations are given by $[\hat{r},\hat{r}^{T}]=i\Omega ,$ where $\Omega
=\oplus _{j=1}^{m}\Omega _{j}$ is the symplectic matrix with $\Omega
_{j}=\left(
\begin{array}{cc}
0 & 1 \\
-1 & 0%
\end{array}%
\right) .$ For convenience, we adopt natural units (%
h{\hskip-.2em}\llap{\protect\rule[1.1ex]{.325em}{.1ex}}{\hskip.2em}%
=$k_{B}$=$1$). In phase space, a Gaussian quantum state $\hat{\rho}$ is
uniquely identified by its characteristic function \cite{35,36},
\begin{equation}
\chi _{G}(r)=\exp \left[ -\frac{1}{4}\tilde{r}^{T}\sigma \tilde{r}+i\tilde{r}%
^{T}d\right] ,  \label{1}
\end{equation}%
where $\tilde{r}=\Omega r$ with $r=(x_{1},p_{1},...,x_{m},p_{m})^{T}$
denoting a vector of 2$m$ real coordinates in phase space, $d=$Tr($\hat{\rho}%
\hat{r}$) represents the first moment, and $\sigma $ is the covariance matrix,
\begin{equation}
\sigma =\text{Tr}\left[ \hat{\rho}\left\{ (\hat{r}-d),(\hat{r}%
-d)^{T}\right\} \right] , \label{2}
\end{equation}%
where the notation $\left\{ \cdot ,\cdot \right\} $ and Tr denote the
anticommutator and the trace of an operator in Hilbert space, respectively.
Thus, the first moment $d$ and the covariance matrix $\sigma $ provide a
complete description of any Gaussian quantum states \cite{35,36}.

In the simultaneous estimation of multiple unknown parameters $\tilde{\theta}
$=$(\theta _{1},...,\theta _{p})^{T},$ the estimation precision is bounded
by the mean square error matrix $\Sigma (\tilde{\theta}),$ which satisfies
the following matrix inequality \cite{37,38,39},
$\Sigma (\tilde{\theta})\geq K^{-1}\geq J^{-1}$,
where $K$ and $J$ denote the classical Fisher information matrix and the
quantum Fisher information matrix, respectively. The entries of the
classical Fisher information matrix are given by
\begin{equation}
K_{uv}=\int dxp(x|\theta )\left( \frac{\partial \ln p(x|\theta )}{\partial
\theta _{u}}\right) \left( \frac{\partial \ln p(x|\theta )}{\partial \theta
_{v}}\right) ,  \label{3}
\end{equation}%
where $p(x|\theta )$\ is the conditional probability distribution. The
quantum Fisher information matrix, which provides a fundamental limit
independent of the specific measurement, has entries given by%
\begin{equation}
J_{uv}=\frac{1}{2}\text{Tr}\left[ \hat{\rho}_{\tilde{\theta}}(\hat{L}_{u}%
\hat{L}_{v}+\hat{L}_{v}\hat{L}_{u})\right] ,  \label{4}
\end{equation}%
where $\hat{L}_{u}$ are the symmetric logarithmic derivative operators
satisfying the Lyapunov equation $\left. \partial \hat{\rho}_{\tilde{\theta}%
}\right/ \partial \theta _{u}=\left. (\hat{L}_{u}\hat{\rho}_{\tilde{\theta}}+%
\hat{\rho}_{\tilde{\theta}}\hat{L}_{u})\right/ 2.$ For a Gaussian quantum
state, uniquely identified by its first moment $d_{\tilde{\theta}}$ 
and the covariance matrix $\sigma _{\tilde{\theta}}$, the classical Fisher
information matrix under a general class of Gaussian measurements
is provided by \cite{40}
\begin{flalign}
K_{uv} =&\left( \frac{\partial d_{\tilde{\theta}}}{\partial \theta _{u}}%
\right) ^{T}\Sigma _{\tilde{\theta}}^{-1}\left( \frac{\partial d_{\tilde{%
\theta}}}{\partial \theta _{v}}\right)
+\frac{1}{2}\text{tr}\left[ \Sigma _{\tilde{\theta}}^{-1}\left( \frac{%
\partial \Sigma _{\tilde{\theta}}}{\partial \theta _{u}}\right) \Sigma _{%
\tilde{\theta}}^{-1}\left( \frac{\partial \Sigma _{\tilde{\theta}}}{\partial
\theta _{v}}\right) \right] ,  \label{5}
\end{flalign}%
where tr denotes the trace of a matrix and $\Sigma _{\tilde{\theta}}=\left.
(\sigma _{\tilde{\theta}}+\sigma _{m})\right/ 2.$ For a single-mode
scenario, $\sigma _{m}=$diag$\left( z,1/z\right) $\ determines the specific
measurement scheme. Specifically, $z=1$\ and\ $z\rightarrow 0$\ represent
heterodyne and homodyne measurements, respectively.

Within the Gaussian framework, the symmetric logarithmic derivative
operators $\hat{L}_{u}$ take the form of Hermitian quadratic operators \cite%
{41,42,43,44},
\begin{equation}
\hat{L}_{u}=L_{u}^{(0)}\text{\^{I}}+\left( L_{u}^{(1)}\right) ^{T}\hat{r}+%
\hat{r}^{T}L_{u}^{(2)}\hat{r},  \label{6}
\end{equation}
where
\begin{flalign}
&L_{u}^{(0)} =-\frac{1}{2}\text{tr}\left[ \sigma _{\tilde{\theta}%
}L_{u}^{(2)}\right] -\left( L_{u}^{(1)}\right) ^{T}d_{\tilde{\theta}}-d_{%
\tilde{\theta}}^{T}L_{u}^{(2)}d_{\tilde{\theta}},  \notag \\
&L_{u}^{(1)} =2\sigma _{\tilde{\theta}}^{-1}\frac{\partial d_{\tilde{\theta}%
}}{\partial \theta _{u}}-2L_{u}^{(2)}d_{\tilde{\theta}},  \notag \\
&\text{vec}[L_{u}^{(2)}] =(\sigma _{\tilde{\theta}}\otimes \sigma _{\tilde{%
\theta}}-\Omega \otimes \Omega )^{-1}\text{vec}\left[ \frac{\partial \sigma
_{\tilde{\theta}}}{\partial \theta _{u}}\right] ,  \label{7}
\end{flalign}%
with vec$[A]$ signifying the vectorization of a matrix $A$.
The resulting quantum Fisher information matrix can be formulated via the
first moment and the covariance matrix \cite{41,42,43,44},
\begin{flalign}
J_{uv} =&\frac{1}{2}\text{vec}\left[ \frac{\partial \sigma _{\tilde{\theta}}%
}{\partial \theta _{u}}\right] ^{T}(\sigma _{\tilde{\theta}}\otimes \sigma _{%
\tilde{\theta}}-\Omega \otimes \Omega )^{-1}\text{vec}\left[ \frac{\partial
\sigma _{\tilde{\theta}}}{\partial \theta _{v}}\right]  \notag \\
&+2\left( \frac{\partial d_{\tilde{\theta}}}{\partial \theta _{u}}\right)
^{T}\sigma _{\tilde{\theta}}^{-1}\frac{\partial d_{\tilde{\theta}}}{\partial
\theta _{v}}.  \label{8}
\end{flalign}

To facilitate a more mathematically tractable analysis, we employ a real,
positive-definite weight matrix $W$ to establish the scalar inequality \cite%
{37,38,39},
\begin{equation}
\text{tr[}W\Sigma (\tilde{\theta})\text{]}\geq C\geq Q,  \label{9}
\end{equation}%
where $C\equiv $tr[$WK^{-1}$] and $Q\equiv $tr[$WJ^{-1}$] are the CRB and
the QCRB, respectively. Throughout this study, we set $W=I_{p}$\ as the
identity matrix, thereby focusing the scalar bound on the total mean square
error of the estimated parameters \cite{40}. This assignment is standard in
unbiased scenarios where each parameter carries equivalent significance a
priori \cite{40}. It is essential to recognize that while the CRB can be
asymptotically reached by a suitable efficient estimator in the large-sample
limit, the QCRB is generally not tight in the multiparameter regime,
although it is always tight for the single-parameter case \cite{45,46,47,48}. 

The attainability of the QCRB is guaranteed if and only if the symmetric
logarithmic derivative operators are mutually compatible, i.e., $[\hat{L}%
_{u},\hat{L}_{v}]=0$\ for all $u$\ and $v$\ \cite{49,50,51,52}. Under such
compatibility, a common basis of eigenstates provides an optimal
measurement basis to saturate the QCRB via single-copy measurements.
However, there are some special quantum states for which $[\hat{L}_{u},%
\hat{L}_{v}]\neq 0$, and the mean Uhlmann curvature matrix $D$ given by the entries
\begin{equation}
D_{uv}=-\frac{i}{2}\text{Tr}\left[ \hat{\rho}_{\tilde{\theta}}[\hat{L}_{u},%
\hat{L}_{v}]\right]\label{10}
\end{equation}%
is a zero matrix \cite{32,53,54,55,56}. In this situation, the QCRB is
asymptotically reachable via the implementation of collective measurements
and the coincides with the Holevo Cram\'{e}r-Rao bound \cite%
{57,58,59,60,61}. Specifically, within the Gaussian regime, the commutation
relation $[\hat{L}_{u},\hat{L}_{v}]$ and the mean Uhlmann curvature matrix
$D$ respectively admit the following analytical
representations \cite{41,42,43,44},
\begin{flalign}
&\lbrack \hat{L}_{u},\hat{L}_{v}] =i\biggl[\left( L_{u}^{(1)}\right) ^{T}\Omega
L_{v}^{(1)}+2\left( L_{u}^{(1)}\right) ^{T}\Omega L_{v}^{(2)}\hat{r}  \notag\\
&\qquad+2\hat{r}^{T}L_{u}^{(2)}\Omega L_{v}^{(1)}+2\hat{r}^{T}L_{u}^{(2)}\Omega
L_{v}^{(2)}\hat{r}-2\hat{r}^{T}L_{v}^{(2)}\Omega L_{u}^{(2)}\hat{r} \biggl],  \notag \\
&D_{uv} =\text{vec}[L_{u}^{(2)}]^{T}\left( \sigma _{\tilde{\theta}}\otimes
\Omega \right) \text{vec}[L_{v}^{(2)}] \notag \\
&\qquad+2\left( \frac{\partial d_{\tilde{\theta}}}{\partial \theta _{u}}\right)
^{T}\sigma _{\tilde{\theta}}^{-1}\Omega \sigma _{\tilde{\theta}}^{-1}\frac{%
\partial d_{\tilde{\theta}}}{\partial \theta _{v}}.  \label{11}
\end{flalign}%
Notably, these two benchmarks coincide for parameters encoded solely in the
first moments. However, the full commutativity of symmetric logarithmic
derivative operators constitutes a stricter condition whenever the
parameters impact the covariance matrix \cite{43}.

\section{Simultaneous estimation of relative phase and coherence}\label{section3}

We investigate the simultaneous estimation of the relative
phase and the coherence by adopting two astronomical
interferometry schemes. 
As illustrated in Fig.~1(a), the first scheme is the direct
interferometry scheme, where two modes, $\hat{a}_{1}$ and $\hat{a}_{2}$, of
the stellar state are collected by separate telescopes (A and B) and
physically interfered \cite{24,27}.Accordingly, the incoming stellar state $%
\hat{\rho}_{s}$ is described as a  correlated thermal state, characterized by its first moment $d_{s} =(0\,0\,0\,0)^{\text{T}}$ and the covariance matrix 
\begin{flalign}
\sigma_{s}=\left(
\begin{array}{cccc}
\epsilon +1 & 0 & \gamma \epsilon \cos \phi & -\gamma \epsilon \sin \phi \\
0 & \epsilon +1 & \gamma \epsilon \sin \phi & \gamma \epsilon \cos \phi \\
\gamma \epsilon \cos \phi & \gamma \epsilon \sin \phi & \epsilon +1 & 0 \\
-\gamma \epsilon \sin \phi & \gamma \epsilon \cos \phi & 0 & \epsilon +1%
\end{array}
\right),  \label{12}
\end{flalign}
where $\epsilon $ is the total mean photon number across the two spatial
modes, the relative phase $\phi \in \lbrack 0,2\pi )$ relates to the
location of the sources, and $\gamma \in \lbrack 0,1]$ represents the degree
of coherence, which is proportional to the Fourier transform of the source
intensity distribution (the object's spatial profile) via the van
Cittert--Zernike theorem \cite{8}. In this framework, $\gamma =1$
corresponds to a point source, whereas a decreased $\gamma$ signifies an
increase in the object's spatial extent.
\begin{figure}[H]
\label{Fig1} \centering\includegraphics[width=0.9\columnwidth]{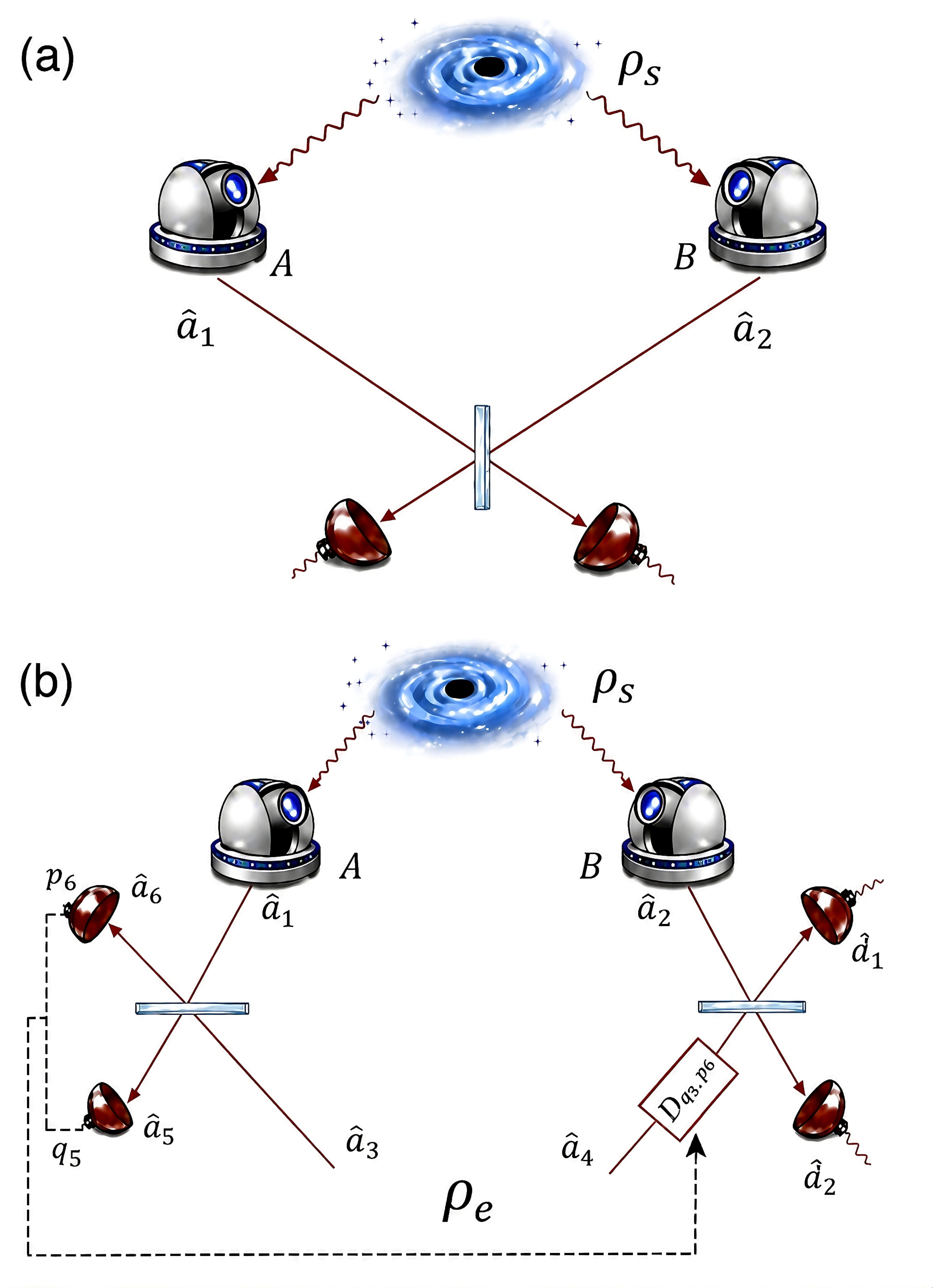}%
\caption{(Color online) Comparison of two schemes for the simultaneous estimation of
the relative phase and the degree of coherence in astronomical interferometry. (a)
Direct interferometry scheme: two modes, $\hat{a}_{1}$ and $\hat{a}_{2}$, of
the stellar state $\hat{\protect\rho}_{s}$ are collected by separated
telescopes (A and B) and physically interfered. Each mode undergoes
transmission loss characterized by an identical transmissivity $T.$ (b)
Continuous-variable quantum teleportation scheme: a two-mode squeezed vacuum
$\hat{\protect\rho}_{e}$ with modes $\hat{a}_{3}$ and $\hat{a}_{4}$ is
distributed via lossy channels of transmissivity\ $T.$ At\ telescope A,
modes $\hat{a}_{1}$ and $\hat{a}_{3}$ are combined on a beam splitter, and
homodyne outcomes $q_{5}$ and $p_{6}$ are sent via a classical channel
represented by the dashed line. A conditional displacement operation $\hat{D}%
_{q_{5},p_{6}}$ is applied to mode $\hat{a}_{4}$ to teleport the state from
telescope A, which is then mixed with local mode $\hat{a}_{2}$ at telescope
B.}
\end{figure}

Assume that each mode undergoes transmission loss characterized by an identical
transmissivity $T$. The first moment and covariance matrix of the stellar
state are transformed to
\begin{flalign}
d_s^{\prime}=d_s,~~
\sigma _{s}^{\prime } =X\sigma _{s}X^{T}+Y, \label{13}
\end{flalign}
where $X=TI_{4}$ and $Y=(1-T^{2})I_{4}$ with $I_{4}$ being the $4\times 4$
identity matrix. 

The second scheme is the continuous-variable quantum teleportation
scheme shown in Fig. 1(b) \cite{25}. In this configuration, the stellar
state is captured by telescopes A and B, corresponding to modes $\hat{a}_{1}$
and $\hat{a}_{2},$ respectively. Concurrently, a two-mode squeezed vacuum
state, $\hat{\rho}_{e}=\left\vert r\right\rangle \left\langle r\right\vert$ 
with $\left\vert r\right\rangle =\exp [r(\hat{a}_{3}^{\dagger }\hat{a}%
_{4}^{\dagger }-\hat{a}_{3}\hat{a}_{4})]\left\vert 00\right\rangle $ and $r$
being the squeezing parameter, serves as the entanglement resource and 
is distributed to the two telescopes via transmission loss channels,
each with an identical transmissivity $T$ corresponding to modes $\hat{a}_{3}$ 
and $\hat{a}_{4}.$ At telescope A, the received stellar mode $\hat{a}_{1}$ 
and the entanglement mode $\hat{a}_{3}$ are combined on a beam
splitter. Homodyne detection is then performed on the two output ports,
yielding the measurement outcomes $q_{5}$ and $p_{6}$. These results are
transmitted to telescope B via a classical communication channel 
denoted by the dashed line. At telescope B, a conditional displacement
operation $\hat{D}_{q_{5},p_{6}}$ is applied to the local entanglement mode 
$\hat{a}_{4}$ based on the received classical information, completing the
quantum teleportation of the stellar state from telescope A to B. Then, the
teleported mode is mixed with telescope B's local stellar mode $\hat{a}_{2}$
on a beam splitter. The first moment and the covariance matrix of the
resulting teleported state are given by \cite{25}
\begin{flalign}
d_{t} =d_{s},~~
\sigma _{t} =\sigma _{s}+4e^{-2\lambda }\left(
\begin{array}{cccc}
1 & 0 & 0 & 0 \\
0 & 1 & 0 & 0 \\
0 & 0 & 0 & 0 \\
0 & 0 & 0 & 0%
\end{array}\right),  \label{14}
\end{flalign}
where $\lambda $ is the effective squeezing parameter and $e^{-2\lambda
}=T^{2}e^{-2r}+1-T^{2}.$ Specifically, $T=1$ represents the lossless
scenario. In the limit of infinite squeezing, $r\rightarrow \infty $ and 
$T=1,$ the teleportation process becomes ideal and the covariance
matrix of the teleported state reduces to that of the stellar state,
$\sigma _{t}=\sigma _{s}.$

Next, we analyze the simultaneous estimation of the relative phase $\phi $
and the degree of coherence $\gamma$ for both direct interferometry and
continuous-variable quantum teleportation schemes.
Concerning the simultaneous estimation, a problem raised in Ref. \cite{24} is that, 
although the quantum Fisher information matrix for the relative phase $\phi$ and 
the degree of coherence $\gamma$ is diagonal, the corresponding symmetric logarithmic 
derivative operators do not commute, and hence the individually optimal measurements 
cannot be implemented simultaneously in general. Nevertheless, according to 
Eqs. (\ref{11}), (\ref{13}), and (\ref{14}), we find that the symmetric logarithmic 
derivative operators $\hat{L}_{\phi }$ and $\hat{L}_{\gamma }$ are non-commutative, 
i.e., $\left[ \hat{L}_{\phi },\hat{L}_{\gamma }\right] \neq 0$ for both aforementioned schemes. 
This non-commutativity implies that a common set of eigenstates for $\hat{L}_{\phi }$ and 
$\hat{L}_{\gamma }$ cannot be established, so the single-parameter optimal measurements 
for $\phi$ and $\gamma$ are not jointly optimal at the level of a common symmetric logarithmic derivative operator eigenbasis. 
Moreover,by substituting Eqs. (\ref{13}) and (\ref{14}) into Eq. (\ref{11}), we further find that the mean Uhlmann curvature matrix is a zero matrix in these cases, 
which indicates that the QCRB remains an asymptotically tight precision limit.
Based on this result, we then substitute Eqs. (\ref{13}) and (\ref{14}) into Eqs. (\ref{8}) and (\ref{9}) to derive the QCRBs for the direct interferometry and continuous-variable quantum teleportation schemes, respectively,
\begin{figure}[H]
\label{Fig2} \centering\includegraphics[width=1\columnwidth]{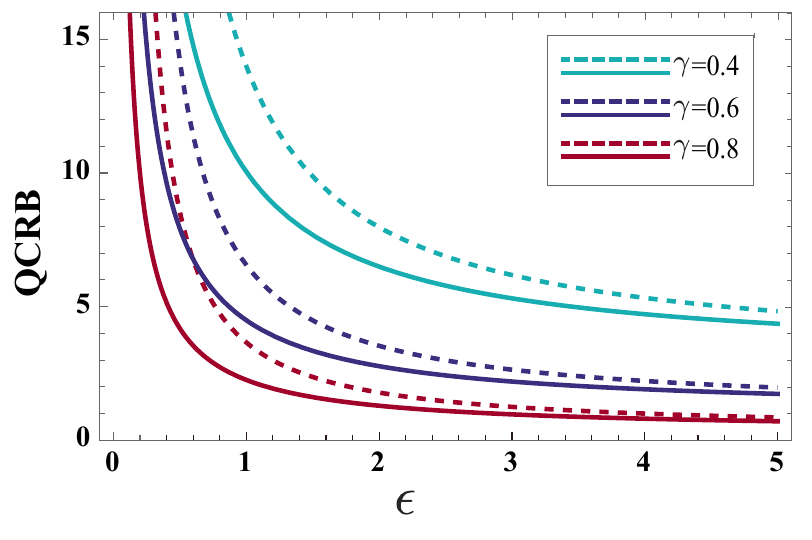}%
\caption{(Color online) The QCRB for the direct interferometry and continuous-variable
quantum teleportation schemes as a function of the total mean photon number $%
\protect \epsilon $\ with $T=1$ and $r=1$. The solid and dashed lines
correspond to the direct interferometry and continuous-variable quantum
teleportation schemes, respectively.}
\end{figure}

\noindent 
\begin{flalign}
Q_{d} =\frac{\Delta _{1}+\Delta _{2}}{\Delta _{3}},~~
Q_{t} =\Delta _{4}+\frac{\Delta _{5}\Delta _{6}\Delta _{7}}{\Delta _{8}},
\label{15}
\end{flalign}
where $Q_{d}$ and $Q_{t}$ correspond to the QCRB for the direct
interferometry and continuous-variable quantum teleportation schemes,
respectively, and $\Delta _{m},$ $m\in \{1,2,3,4,5,6,7,8\}$ are given in
Appendix.

In general, a smaller value of QCRB corresponds to higher estimation
precision. We first consider the lossless scenario ($T=1$) to compare the
ultimate precision limits of the two schemes. Notably, the
continuous-variable quantum teleportation scheme is found to asymptotically
approach the direct interferometry scheme in the limit of infinite squeezing
regime, i.e., $Q_{d}=\lim_{r\rightarrow \infty }Q_{t}$. For cases involving
finite squeezing, based on Eq. (\ref{15}), we plot the QCRB for these two
schemes as a function of the total mean photon number $\epsilon $ for $r=1$
and $\gamma =0.4,$ $0.6,$ $0.8$, see Fig. 2. The results demonstrate
that the direct interferometry scheme consistently yields a lower QCRB,
indicating superior estimation precision compared to the continuous-variable
quantum teleportation scheme. For both schemes, the estimation precision
improves monotonically as $\epsilon $ increases. Furthermore, for a fixed 
$\epsilon $, higher values of $\gamma $ lead to a significant reduction in
QCRB, thereby enhancing the estimation performance of both schemes.

Subsequently, we examine the conditions under which the CRB for Gaussian
measurements saturates the QCRB in both the direct interferometry and
continuous-variable quantum teleportation schemes. Although analytical
expressions for the CRB can, in principle, be derived from Eqs. (\ref{5}), (\ref{9}),
(\ref{13}) and (\ref{14}), the analytical expressions are prohibitively
complex and are thus omitted here. Instead, we resort to a numerical
comparison of the difference $f(z)$ between the CRB $C(z)$ and the QCRB $Q$
for the two schemes, $f(z)=C(z)-Q$. 
\begin{figure}[H]
\label{Fig3} \centering\includegraphics[width=1\columnwidth]{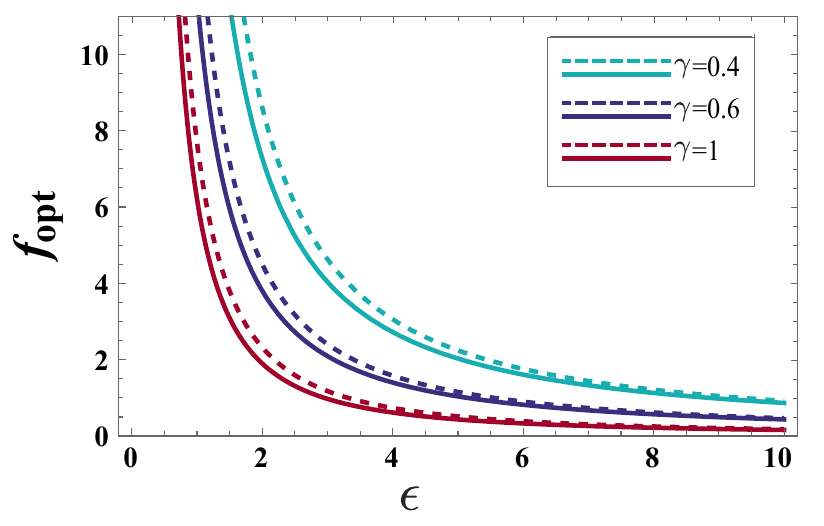}
\caption{(Color online) The optimized difference $f_{\text{opt}}$ for the direct
interferometry and continuous-variable quantum teleportation schemes as a
function of the total mean photon number $\protect\epsilon$ with $T=1,$ 
$\protect\phi =\protect\pi /2$ and $r=0.5.$ The solid and dashed lines
correspond to the direct interferometry and continuous-variable quantum
teleportation schemes, respectively.}
\end{figure}

\noindent
It is evident that a smaller $f(z)$ indicates that the
CRB approaches the QCRB more closely. $f(z)=0$ signifies that the
chosen Gaussian measurement attains the QCRB. To minimize the gap, we define
the optimized difference,
\begin{flalign}
f_{\text{opt}} =\min\limits_{z}f(z)=\min\limits_{z}\left[ C(z)\right] -Q,  \label{16}
\end{flalign}
which represents the optimized difference between CRB and QCRB by
optimizing over the Gaussian measurement parameters $z$. Here, 
$\min\limits_{z}\left[ C(z)\right] $ denotes the optimized CRB. To illustrate
this, we plot the optimized difference $f_{\text{opt}}$ as a function of the
total mean photon number $\epsilon $ for $\gamma =0.4,$ $0.6,$ $1$ in Fig.
3. It is observed that the optimized difference $f_{\text{opt}}$ decreases
monotonically with the increasing $\epsilon $ across all values of the degree of
coherence $\gamma $, indicating that the selected optimal Gaussian
measurement becomes increasingly optimal in the large-$\epsilon $ regime. In the case of $\gamma =1$, both schemes asymptotically saturate the QCRB for 
sufficiently large $\epsilon$. Conversely, a reduction in $\gamma $
shifts the curves upward, suggesting that a lower degree of coherence
increases the difference between the CRB and the QCRB.
Furthermore, for a
fixed degree of coherence, the direct interferometry scheme consistently
exhibits a smaller $f_{\text{opt}}$ than the continuous-variable quantum
teleportation scheme, demonstrating its higher efficiency in approaching the
theoretical precision limit.

In addition, it is natural to identify the
specific type of the optimal Gaussian measurement required. This involves
finding the optimal parameter $z_{\text{opt}}$ that minimizes $f(z)$. The
behavior of $z_{\text{opt}}$ is displayed in Fig. 4.
In the large-$\epsilon$ regime, for both direct interferometry and
continuous-variable quantum teleportation schemes, $z_{\text{opt}}$ remains uniformly at $1$
across the entire phase range. This indicates that heterodyne measurement is
relative phase and degree of coherence.
\begin{figure}[H]
\label{Fig4} \centering\includegraphics[width=1\columnwidth]{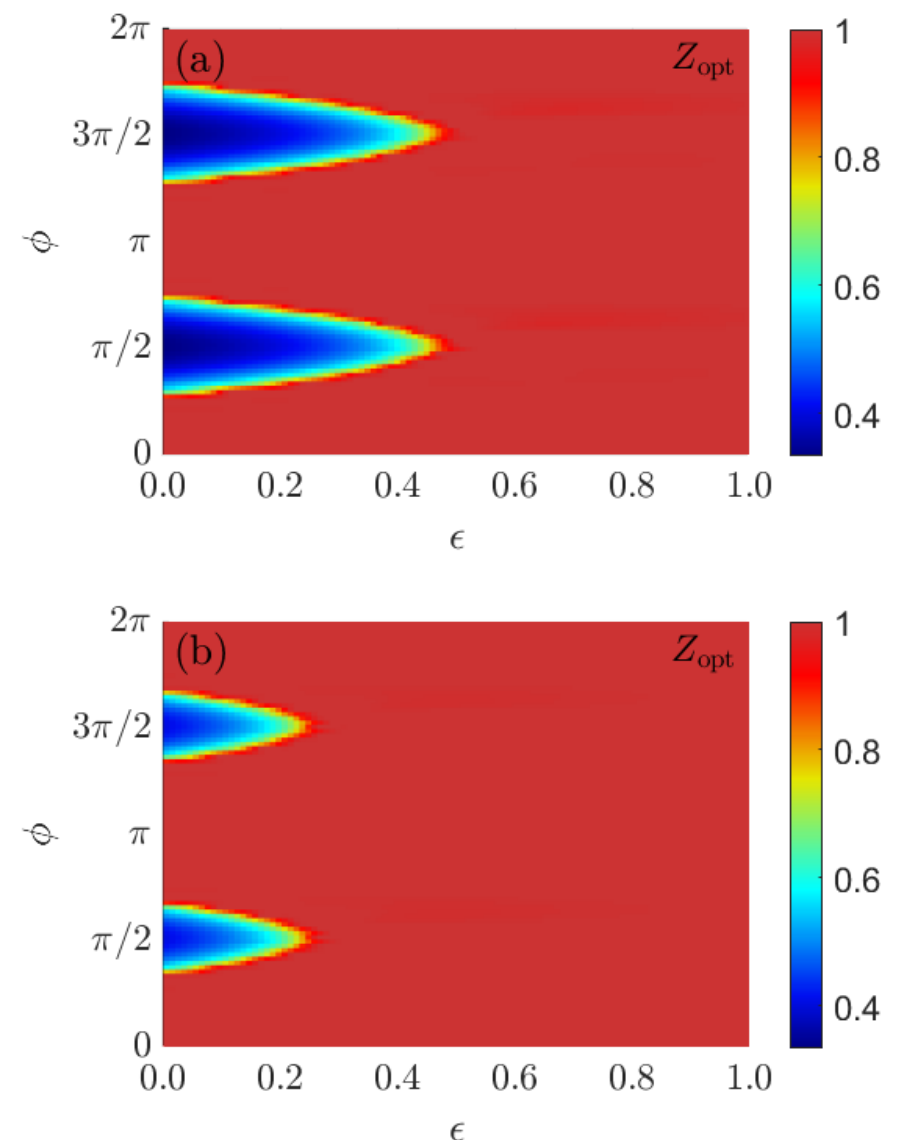}%
\caption{(Color online) The optimal parameter $z_{\text{opt}}$ for the direct
interferometry and continuous-variable quantum teleportation schemes as a
function of the total mean photon number $\protect\epsilon $\ and the
relative phase $\protect\phi$. (a)\ the direct interferometry with $T=1$ and
$\protect\gamma =0.6$. (b) the continuous-variable quantum teleportation
schemes with $T=1,$ $r=1$ and $\protect\gamma =0.6.$}
\end{figure}

\noindent the robustly optimal Gaussian measurement for the simultaneous estimation of
relative phase and degree of coherence. Conversely, in the small-$\epsilon$
regime, the optimality of the measurement scheme becomes highly
phase-sensitive. Specifically, around $\phi =\pi /2,$ $3\pi /2,$ $z_{\text{%
opt}}$ drops below $1$,  suggesting that a general Gaussian measurement is
required to minimize $f(z)$. As $\epsilon$ increases, these
phase-dependent regions gradually shrink and eventually vanish, reflecting a
convergence toward a universal heterodyne measurement strategy. 

Figure. 5(a) depicts the optimized CRB, evaluated at the optimal Gaussian measurement, as
a function of the degree of coherence $\gamma $. It is shown that the CRB
decreases monotonically with an increase in the degree of coherence $\gamma $
for both schemes, confirming that spatial coherence is a fundamental
resource for enhancing estimation precision. Moreover, higher photon numbers
$\epsilon $ lead to a lower CRB, demonstrating that increased source
intensity can effectively compensate for the precision loss caused by low
coherence. This dynamic is further detailed in Fig. 5(b), which plots the
optimized CRB against the total mean photon number $\epsilon $. For
both schemes,
\begin{figure}[H]
\label{Fig5} \centering\includegraphics[width=1\columnwidth]{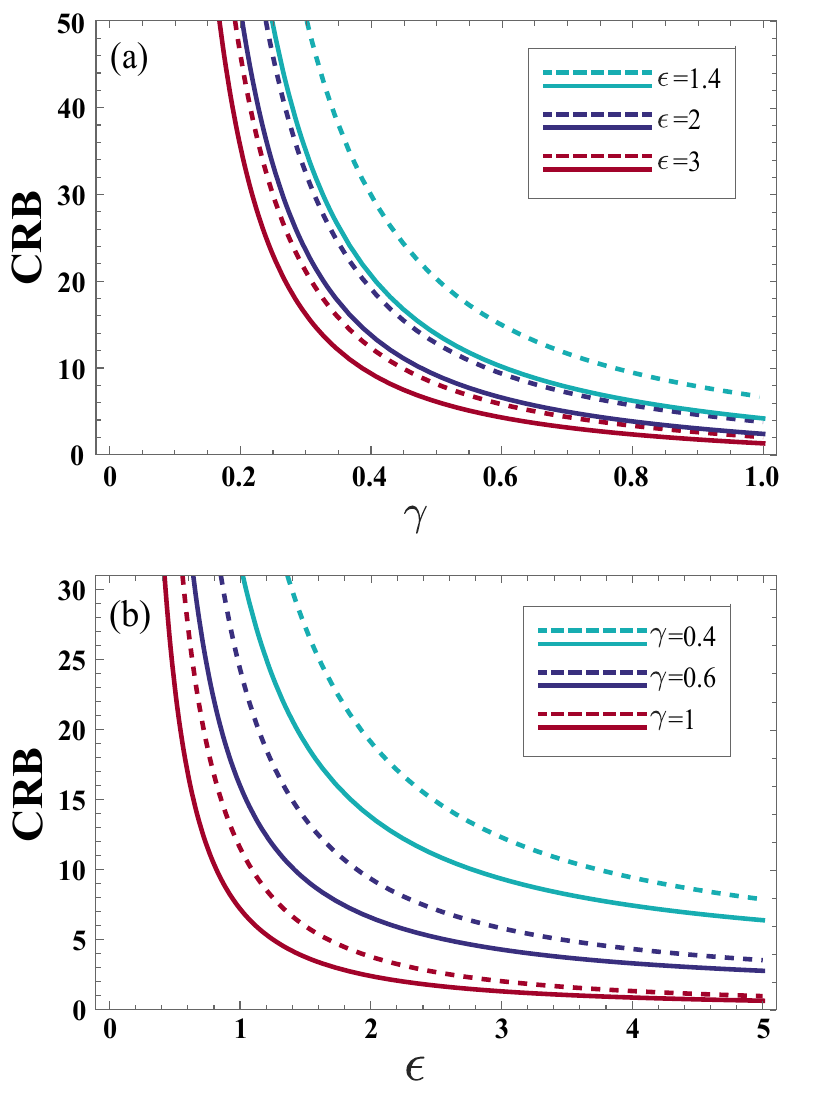}%
\caption{(Color online) The optimized CRB for the direct interferometry and
continuous-variable quantum teleportation schemes as a function of (a) the
degree of coherence $\protect\gamma $ and of (b) the total mean photon
number $\protect\epsilon $ with $T=1$, $\protect\phi =\protect\pi /2$ and 
$r=0.5.$ The solid and dashed lines correspond to the direct interferometry
and continuous-variable quantum teleportation schemes, respectively.}
\end{figure}

\noindent  the CRB decays rapidly with increasing $\epsilon $,  demonstrating
that higher source intensity is a primary driver for enhancing estimation
precision. For a fixed degree of coherence, the direct interferometry scheme
consistently yields a lower CRB than the continuous-variable quantum
teleportation scheme, particularly in the low-photon regime. Furthermore, a
reduction in the degree of coherence (from $\gamma $ $=1$ 
to $\gamma $ $=0.4$) results in an upward shift of the CRB curves.
This trend highlights that
while increasing the mean photon number improves the bound, the overall
performance remains fundamentally constrained by the initial coherence of
the system.

Next, we evaluate the impact of transmission loss on the estimation
precision for both direct interferometry and continuous-variable quantum
teleportation schemes. The transmissivity $T$ associated with these schemes
is given by \cite{25}, $T=10^{-{\alpha L}/{20}}$,
where $\alpha $ is the transmission loss coefficient and $L$ the
baseline length. For our numerical analysis, we assume a typical
transmission loss of $\alpha =$ $0.2$ dB/Km, which corresponds to the standard
optical fibers operating at
\begin{figure}[H]
\label{Fig6} \centering\includegraphics[width=1\columnwidth]{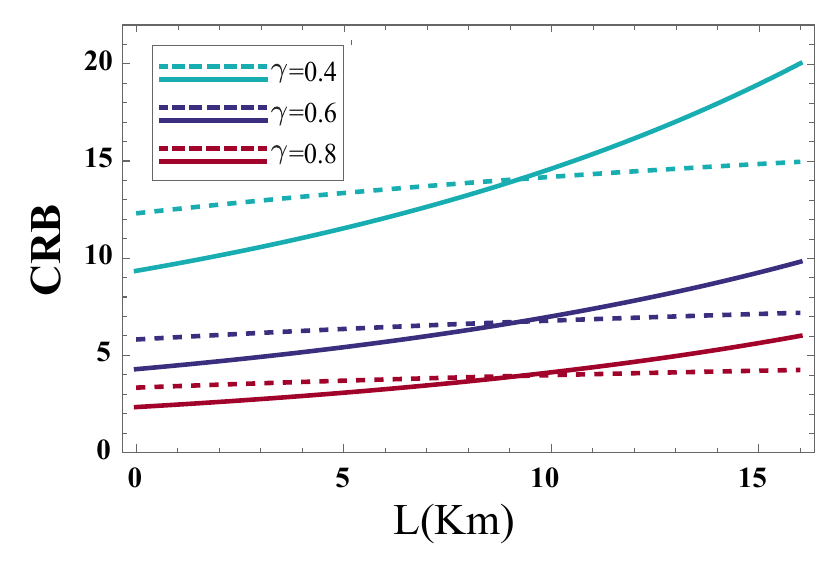}%
\caption{(Color online) The optimized CRB for the direct interferometry and
continuous-variable quantum teleportation schemes as a function of the
baseline length $L$\ with $\protect\epsilon =3,$ $\protect\phi =\protect\pi %
/2$, $r=0.5$ and the transmission loss coefficient $\protect\alpha =0.2$
dB/Km. The solid and dashed lines correspond to the direct interferometry
and continuous-variable quantum teleportation schemes, respectively.}
\end{figure}
\noindent $1550$ nm. Figure. 6 illustrates the optimized CRB
based on the optimal Gaussian measurement for both schemes as a function of
the baseline length $L$ for three different degrees of coherence $\gamma
=0.4,$ $0.6,$ $0.8$. As expected, the CRB increases with $L$ in all cases
due to the accumulation of transmission loss. Notably, a distinct
crossover behavior emerges between the two measurement strategies. In the
short-baseline regime, the direct interferometry scheme yields a lower CRB,
indicating superior estimation precision. However, as the baseline length
increases, the precision of the direct scheme degrades much more rapidly
than that of the continuous-variable quantum teleportation scheme.
Consequently, at longer baselines, the continuous-variable quantum
teleportation scheme outperforms the direct approach, maintaining a more
stable and lower CRB. Furthermore, a reduction in the degree of coherence
shifts all curves upward and accentuates the rapid degradation of the direct
scheme, emphasizing the robustness of the continuous-variable quantum
teleportation scheme for long-distance transmission in lossy environments.

\section{Conclusions}\label{section4}

In summary, we have presented a detailed analysis of the simultaneous estimation of
the relative phase and the degree of coherence in astronomical interferometry
for both the direct interferometry scheme and the continuous-variable quantum
teleportation scheme. In these two schemes, the symmetric logarithmic
derivative operators for relative phase and degree of coherence are
non-commuting, and the mean Uhlmann curvature matrix is a zero matrix,
indicating that the QCRB represents an asymptotically tight precision limit.
We have shown that the direct interferometry scheme
consistently yields a lower QCRB, offering superior estimation precision
over the continuous-variable quantum teleportation scheme. For both schemes,
our numerical results show that the QCRB decreases as both the total mean photon 
number $\epsilon $ and the degree of coherence $\gamma $ increase. 
We have further examined the conditions under which the CRB for Gaussian measurements 
saturates the QCRB. In the case of $\gamma =1$, both schemes asymptotically saturate the QCRB as
the total mean photon number becomes sufficiently large. Heterodyne
measurement is near-optimal for both direct interferometry and
continuous-variable quantum teleportation schemes in large-$\epsilon$
regime. Moreover, we have evaluated the impact of the transmission loss on estimation
precision. In the short-baseline regime, the direct interferometry scheme
provides a lower CRB, indicating superior precision. Conversely, at longer
baselines, the continuous-variable quantum teleportation scheme outperforms
the direct interferometry scheme by maintaining a lower and more stable CRB.

\Acknowledgements{We sincerely thank Marco G. Genoni and Francesco Albarelli for useful
discussions. This work was supported by the National
Key Research and Development Program of China (Grants No. 2021YFA1400900,
No. 2021YFA0718300, and No. 2021YFA1400243), NSFC (Grants No. 12074105, No. 12104135, No. 12404377, and No. 61835013), 
Natural Science Foundation of Henan province (Grant No. 252300421995), and Key Scientific Re search
Projects of Henan Province Higher Education Institu tions (Grant No.
26B140007). Wei Ye is supported by Jiangxi Provincial Natural Science
Foundation (20232BAB211032, 20252BAC240169) and the Scientific Research
Startup \ Foundation (EA202204230).}

\textbf{Appendix: Coefficients in Eq.} (\ref{15})

The corresponding coefficients of the QCRBs for
the direct interferometry and continuous-variable quantum teleportation
schemes in the Eq. (\ref{15}):
\begin{flalign}
\Delta _{1} =&4+4\gamma ^{2}(1-\gamma ^{2})+4\epsilon T^{2}(1+\gamma
^{2}-\gamma ^{4}),  \notag \\
\Delta _{2} =&\epsilon ^{2}T^{4}(1+\gamma ^{2}-3\gamma ^{4}+\gamma ^{6}),
\notag \\
\Delta _{3} =&2\epsilon T^{2}\gamma ^{2}\left[ 2+\epsilon T^{2}(1+\gamma
^{2})\right] , \notag \\
\Delta _{4} =&\frac{4e^{-2\lambda }(1+\epsilon )+\epsilon (2+\epsilon
-\gamma ^{2}\epsilon )}{2\epsilon ^{2}\gamma ^{2}},  \notag \\
\Delta _{5} =&\epsilon (\gamma ^{2}-1)-4e^{-2\lambda },  \notag \\
\Delta _{6} =&(2+\epsilon +\gamma \epsilon )[\epsilon (\gamma -1)-2]  \notag
-4e^{-2\lambda }(2+\epsilon ),  \notag \\
\Delta _{7} =&\epsilon \lbrack \epsilon (\gamma ^{2}-1)-2]-4e^{-2\lambda
}(1+\epsilon ),  \notag \\
\Delta _{8} =&2\epsilon ^{2}[\gamma ^{4}\epsilon ^{3}-16e^{-4\lambda }(2+\epsilon ) \notag \\
&-8e^{-2\lambda }(1+\epsilon )(2+\epsilon )-\epsilon
(2+\epsilon )^{2}].  \nonumber
\end{flalign}

\InterestConflict{
The authors declare no conflicts of interest.}


\InterestConflict{The authors declare that they have no conflict of interest.}




\end{multicols}

\begin{thebibliography}{99}
\bibitem{1} Z. X. Huang, O. Titov, M. K. Schmidt, B. Pope, G. K. Brennen, D.
K. L. Oi, and P. Kok, Adv. Phys. X \textbf{10}, 2597311 (2025).

\bibitem{2} J. D. Monnier,  Rep. Prog.
Phys. \textbf{66}, 789 (2003).

\bibitem{3} P. J. Stas, Y. C. Wei, M. Sirotin, Y. Q. Huan, U. Yazlar, F. A.
Arias, E. Knyazev, G. Baranes, B. Machielse, S. Grandi, D. Riedel, J.
Borregaard, H. Park, M. Lon\v{c}ar, A. Suleymanzade, and M. D. Lukin,
Nature \textbf{651}, 326 (2026).

\bibitem{4} Y. Wang, and E. Chitambar,  Phys. Rev. Lett.
\textbf{134}, 170801 (2025).

\bibitem{5} B. Wang, X. Y. Luo, B. F. Gao, J. L. Liu, C. Y. Wang, Z. Yan, Q. M. Ke, D. Teng, M. Y. Zheng, Y. Cao, J. Li, C. Z. Peng, Q. Zhang, X. H. Bao, and J. W. Pan, Phys. Rev. Lett.  \textbf{136}, 240801 (2026).

\bibitem{6} M. R. Brown, M. Allgaier, V. Thiel, J. D. Monnier, M. G. Raymer, and B. J. Smith, Phys. Rev. Lett. \textbf{131}, 210801 (2023).

\bibitem{7} A. Sajjad, M. R. Grace, and S. Guha, Phys. Rev. Res. \textbf{6}, 013212 (2024).

\bibitem{8} F. Zernike,  Physica \textbf{5%
}, 785 (1938).

\bibitem{9} M. Tsang, Phys. Rev. Lett.
\textbf{107}, 270402 (2011).

\bibitem{10} A. R. Kurek, T. Pi\k{e}ta, T. Stebel, A. Pollo, and A.
Popowicz,  Opt. Lett. \textbf{41}, 1094 (2016).

\bibitem{11} J. Leng, Y. X. Shen, Z. K. Cao, and X. B. Wang,
 Opt. Express
\textbf{34} 1424 (2026).

\bibitem{12} Y. Wang, Y. Zhang, and V. O. Lorenz, Phys. Rev. Lett. \textbf{135}, 113602 (2025).

\bibitem{13} D. Gottesman, T. Jennewein, and S. Croke,  Phys.
Rev. Lett. \textbf{109}, 070503 (2012).

\bibitem{14} E. T. Khabiboulline, J. Borregaard, K. De Greve, and M. D.
Lukin,  Phys. Rev. Lett. \textbf{123}, 070504 (2019).

\bibitem{15} E. T. Khabiboulline, J. Borregaard, K. De Greve, and M. D.
Lukin,  Phys. Rev. A \textbf{100}, 022316 (2019).

\bibitem{16} M. M. Marchese, and P. Kok,  Phys. Rev. Lett. \textbf{130}, 160801 (2023).

\bibitem{17} Y. X. Shen, Z. K. Cao, J. Leng, and X. B. Wang,
 Phys. Rev. Res. \textbf{7}, 043110
(2025).

\bibitem{18} C. Y. Hu, B. Wang, J. D. Zhang, K. X. Wang, H. G. Liu, J. Zhou,
and L. J. Zhang,  Phys. Rev. A
\textbf{112}, 032609 (2025).

\bibitem{19} Y. Zhang, and T. Jennewein,
Phys. Rev. Res. \textbf{7}, 043278 (2025).

\bibitem{20} I. Padilla, A. Sajjad, B. N. Saif, and S. Guha,
 Phys. Rev. Lett. \textbf{136}, 010803
(2026).

\bibitem{21} I. Padilla, A. Sajjad, B. N. Saif, and S. Guha,
 Phys. Rev. A \textbf{113},012608 (2026).

\bibitem{22} Z. X. Huang, G. K. Brennen, and Y. K. Ouyang,  Phys. Rev.Lett. \textbf{129}, 210502 (2022).

\bibitem{23} Z. X. Huang, and C. Lupo,  Phys. Rev. A
\textbf{110}, 052431 (2024).

\bibitem{24} Z. X. Huang, B. Q. Baragiola, N. C. Menicucci, and M. M. Wilde, Phys. Rev. A \textbf{%
109}, 052434 (2024).

\bibitem{25} Y. Wang, Y. Zhang, and V. O. Lorenz,  Phys. Rev. Res. \textbf{7}, 023154
(2025).

\bibitem{26} B. Purvis, R. Lafler, and R. N. Lanning,  New J. Phys. \textbf{26}, 103006 (2024).

\bibitem{27} Y. Wang, and S. Zhou,  Phys. Rev. Lett. \textbf{%
135}, 120201 (2025).

\bibitem{28} R. Czupryniak, J. Steinmetz, P. G. Kwiat, and A. N. Jordan,
 Phys. Rev. A \textbf{108}, 052408 (2023).

\bibitem{29} R. Czupryniak, E. Chitambar, J. Steinmetz, and A. N. Jordan,Phys.Rev. A \textbf{106}, 032424 (2022).

\bibitem{30} S. Modak, and P. Kok,  Phys. Rev. A \textbf{111}, 043701 (2025).

\bibitem{31} S. L. Braunstein and C. M. Caves,  Phys. Rev.Lett. \textbf{72}, 3439 (1994).

\bibitem{32} F. Albarelli, M. Barbieri, M. G. Genoni, and I. Gianani,Phys. Lett. A \textbf{384}, 126311 (2020).

\bibitem{33} A. Carollo, B. Spagnolo, and D. Valenti, Sci.Rep. \textbf{8}, 9852 (2018).

\bibitem{34} R. Demkowicz-Dobrza\'{n}ski, W. G\'{o}ecki, and M. Gut\v{a},J. Phys. A: Math. Theor. \textbf{53},
363001 (2020).

\bibitem{35} S. L. Braunstein and P. van Loock,  Rev. Mod. Phys. \textbf{77}, 513 (2005).

\bibitem{36} C. Weedbrook, S. Pirandola, R. Garc\'{\i}-Patr\'{o}n, N. J.
Cerf, T. C. Ralph, J. H. Shapiro, and S. Lloyd,  Rev. Mod. Phys. \textbf{84}, 621
(2012).

\bibitem{37} J. S. Sidhu and P. Kok,  AVS Quantum
Sci. \textbf{2}, 014701 (2020).

\bibitem{38} V. Montenegro, C. Mukhopadhyay, R. Yousefjani, S. Sarkar, U.
Mishra, M. G. A. Paris, and A. Bayat,  Phys. Rep.
\textbf{1134}, 1-62 (2025).

\bibitem{39} J.~Liu, H. D.~Yuan, X. M.~Lu, and~X. G.~Wang,  J. Phys. A: Math. Theor. \textbf{53},
023001 (2020).

\bibitem{40} G. Bressanini, M. G Genoni, M. S. Kim, and M. G. A. Paris,
 J. Phys. A: Math. Theor. \textbf{57}%
, 315305 (2024).

\bibitem{41} R. Nichols, P. Liuzzo-Scorpo, P. A. Knott, and G. Adesso,
 Phys. Rev. A \textbf{98}, 012114 (2018).

\bibitem{42} Y. Gao and H. Lee,  Eur. Phys. J. D
\textbf{68}, 347 (2014).

\bibitem{43} S. K. Chang, M. G. Genoni, and F. Albarelli,
Commun. Phys. \textbf{9}, 126 (2026).

\bibitem{44} D. \v{S}afr\'{a}ek,  J. Phys. A: Math. Theor. \textbf{52},035304 (2019).

\bibitem{45} H. Chen, Y. Chen, and H. Yuan,  Phys. Rev. A \textbf{%
105}, 062442 (2022).

\bibitem{46} F. Albarelli, J. F. Friel, and A. Datta,  Phys. Rev. Lett. \textbf{123}, 200503
(2019).

\bibitem{47} S. Imai, J. Yang, and L. Pezz\`{e},  Phys. Rev. Lett. \textbf{136}, 150801
(2026).

\bibitem{48} S. S. Zhou, and S. Chen,  PRX
Quantum \textbf{7}, 010314 (2026).

\bibitem{49} L. O. Conlon, J. Suzuki, P. K. Lam, and S. M. Assad,
 Phys. Lett. A \textbf{542}, 130445 (2025).

\bibitem{50} M. Hayashi and Y. Ouyang,  npj Quantum Inf. \textbf{10}, 111 (2024).

\bibitem{51} S. K. Yung, L. O. Conlon, J. Zhao, P. K. Lam, and S. M. Assad,
 Phys. Rev. Res. \textbf{6}, 033315 (2024).

\bibitem{52} F. Albarelli, and R. Demkowicz-Dobrza\'{n}ski,
 Phys. Rev. X \textbf{12}, 011039 (2022).

\bibitem{53} L. O. Conlon, J. Suzuki, P. K. Lam, and S. M. Assad, npj Quantum Inf. \textbf{7}, 110 (2021).

\bibitem{54} X. M. Lu and X. G. Wang,  Phys. Rev. Lett. \textbf{126}, 120503 (2021).

\bibitem{55} J. Yang, S. Pang, Y. Zhou, and A. N. Jordan,  Phys. Rev. A \textbf{100}, 032104 (2019).

\bibitem{56} M. G. Genoni, M. G. A. Paris, G. Adesso, H. Nha, P. L. Knight,
and M. S. Kim,  Phys. Rev. A \textbf{87}, 012107 (2013).

\bibitem{57} J. Suzuki,
J. Math. Phys. \textbf{57}, 042201 (2016).

\bibitem{58} J. Suzuki,  Entropy \textbf{21}, 703 (2019).

\bibitem{59} L. Pezz\`{e}, M. A. Ciampini, N. Spagnolo, P. C. Humphreys, A.
Datta, I. A. Walmsley, M. Barbieri, F. Sciarrino, and A. Smerzi, Phys. Rev. Lett. \textbf{119},
130504 (2017).

\bibitem{60} Y. Yang, G. Chiribella, and M. Hayashi,  Commun. Math. Phys. \textbf{368}, 223
(2019).

\bibitem{61} S. Ragy, M. Jarzyna, and R. Demkowicz-Dobrza\'{n}ski, Phys. Rev. A \textbf{94}, 052108 (2016).


\end{thebibliography}
\end{document}